\documentclass[twocolumn,showpacs,showkeys,amsfonts,prb,amsmath,amssymb,floatfix,footinbib]{revtex4}

\usepackage[pdftex]{color,graphicx}
\usepackage[tight]{subfigure}
\usepackage{dcolumn}
\usepackage{bm}
\usepackage{xspace}

\newcommand{\mub}{\ensuremath{\mu_\text{B}}\xspace}
\newcommand{\mom}[2]{\ensuremath{#1\ \mub/\mathrm{#2}}} 
\newcommand{\nmom}[2]{\ensuremath{\sim \mom{#1}{#2}}}
\newcommand{\RmtKmax}{\ensuremath{R_{\mathrm{MT}}*K_{\mathrm{max}} }\xspace }

\newcommand{\units}[1]{\ensuremath{\ \mathrm{#1}}\xspace}
\newcommand{\ud}{\ensuremath{\mathrm{d}} }

\newcommand{\dz}{d$_{2z^2 - x^2 - y^2}$\xspace}
\newcommand{\dxyxy}{d$_{x^2 - y^2 + xy}$\xspace}
\newcommand{\dxzyz}{d$_{xz + yz}$\xspace}

\newcommand{\Umat}{ \ensuremath{ U_{m m\prime}}\xspace}
\newcommand{\Jmat}{\ensuremath{ J_{m m\prime}}\xspace}
 
\newif{\ifDMused}
\DMusedtrue

\newif{\ifAMFused}
\AMFusedfalse
\newcommand{\AMF}{%
	\ifAMFused 
		AMF 
	\else 
		Around Mean Field (AMF) 
	\fi
	\AMFusedtrue
	\xspace
}

\newcommand{\figref}[1]{Fig.~\ref{#1}}
\newcommand{\eqnref}[1]{Eqn.~(\ref{#1})}
\newcommand{\tblref}[1]{Tbl.~\ref{#1}}
\newcommand{\linecite}[1]{Ref.\,\onlinecite{#1}\xspace}

\newcounter{TODOCount}

\makeatletter 
	\newcommand \listoftodos{
		\clearpage
		\onecolumngrid
		\section*{Todo list} 
		\@starttoc{tdo}
		}
  	\newcommand \l@todo[2]{\par\noindent \textit{#2}, \parbox{10cm}{#1}\par} 
\makeatother

\begin{document}

\title{The Role of Correlations in the Magnetic Moment of MnSi}
\author{Robert D. Collyer}
\author{Dana A. Browne}
\affiliation{Louisiana State University, Baton Rouge, Louisiana, USA}

\date{\today}

\begin{abstract}
We explored the anomalously low moment in MnSi found by experiment (\nmom{0.4}{Mn}) vs the moment predicted by density functional theory (\nmom{1.0}{Mn}).  With the addition of a Hubbard-U correction, we found several solutions with lower moments.  These lower moment solutions show an unusual magnetic order and a magnetic quadrupole moment.  The behavior of the moment under pressure does not follow the experimental trend.
\end{abstract}


\maketitle

\section{Introduction}

A number of transition metal binary compounds can crystallize in the B20 structure and exhibit a wide variety of behaviors.\cite{Schob:a04144,mnge..coge,vanderMarel1998138,Wernick19721431,ishikawa76,PhysRevB.35.2267}  In particular, MnSi is a magnetic metal with an ordered moment $\mu = 0.4 \units{\mub/Mn}$, and the order manifests as a long period ($\sim 180 \units{\AA}$) spin helix along the [111] direction due to the lack of inversion symmetry.\cite{ishikawa76}  Above the relatively small T$_\text{c}$ of $29 \units{K}$, MnSi exhibits Curie-Weiss like behavior with a moment of $\mu_{CW} = 2.2 \units{\mub/Mn}$ which indicates that the magnetism is due to itinerant electrons.\cite{corti:115111}  More recent attention has focused on its metamagnetic transition with the application of $\sim 1.46 \units{GPa}$, above which MnSi exhibits non-Fermi liquid behavior.\cite{MnSiPressure.exp.97,MnSiPressure.exp.97.2, MnSiPressure.exp.00,non-fermi:01}  Similarly, FeGe is an itinerant helimagnet with an ordered moment $\mu = 1.0 \units{\mub/Fe}$, a longer period ($\sim 700 \units{\AA}$), and a high temperature phase where the helix lies along the [100] direction which changes to the [111] direction in a transition with a large temperature hysteresis ($\Delta T \simeq 30 \units{K}$).\cite{lbf89}  Additionally, the magnet order is likewise destroyed with the application of $20 \units{GPa}$ of pressure, but the high pressure phase shows a residual magnetization that is attributed to disorder.\cite{pedrazzini2006msc}  In contrast, FeSi is a small gap semiconductor that undergoes an unusual metal to insulator transition.\cite{jaccarino67,schlesinger93,Saitoh1995307}  There is an ongoing debate on whether its properties are due to Kondo-like\cite{fu:doniach:95,PhysRevB.58.15483} or band-like interactions,\cite{PhysRevLett.76.1735,prl.101.046406} among others.\cite{jarlborg:205105} 

Previous DFT calculations have had mixed results predicting the behaviors of the B20 compounds.  FeSi is correctly predicted to be a nonmagnetic semiconductor with a gap that is comparable to experiment.\cite{mattheiss93.fesi}  Similarly, FeGe is predicted to have a FM moment of \nmom{1.0}{Fe} at the experimental volume.\cite{fege03,fesige04:3,fesige04:2,pedrazzini2006msc}  However, MnSi has been predicted to have a similar moment between \nmom{0.9}{Mn} and \mom{1.2}{Mn} when calculated using the experimental lattice constant using either an LMTO or LAPW method.\cite{Lerch1994321,MnSiPressure.dft.99,corti:115111,jeongpickett04}  At the LDA minimum volume, MnSi is predicted to be either paramagnetic\cite{MnSiPressure.dft.99} or ferromagnetic with a moment in the range $0.75$ to \mom{0.79}{Mn}.\cite{corti:115111,Lerch1994321}  Calculations for both MnSi and FeGe have shown their metamagnetic transitions under pressure, but the predicted moment collapse requires at least twice the experimental pressure when starting from the experimental volumes.\cite{MnSiPressure.dft.99,neef:035122}  The noted similarities between the predicted density of states of the B20 monosilicides\cite{mena_thesis} suggests that a fixed spin moment calculation may yield useful results,\cite{jeongpickett04} but this solution is unsatisfactory as it adds an additional chemical potential to the calculation while not addressing the underlying cause of the discrepancy.  However, the narrowness of the Mn 3d -- Si 3p bands compared to the Fe 3d -- Ge 4p bands and the non-negligible electronic interactions suggest that additional correlations are needed.\cite{carbone:085114}

In this paper, we examine the effect of adding a Hubbard-$U$ interaction to our DFT calculations in the form of an {LDA+U} functional.  In general, the Hubbard-U is a fixed property of the system.  But, {LDA+U} lacks dynamic screening, so the calculated value usually differs from the measured value.\cite{PhysRevLett.76.1735}  Considering this and the inherent variability in DFT exhibited by the prior MnSi calculations, we chose to vary $U$ over a range including the experimental value\cite{carbone:085114} and a value calculated for MnO.\cite{mno:08}  This allowed us to fit the experimental data better, and for values of $U$ similar to the MnO value, we found a ground state with a moment comparable to experiment.\cite{mno:08}

Depending on the value of $U$ chosen, our calculations show two different ground states connected by a complex transition region.  The low-$U$ ground states retain a moment of \mom{1}{Mn} while showing decreased magneto-volume coupling and becoming half-metallic as $U$ is increased.  However, the high-$U$ ground states are very different exhibiting a marked reduction in moment, a significant rearrangement of the spin density, and a non-zero magnetic quadrupole moment around the manganese.  

Section \ref{sec:method} details the specific DFT method we used.  Section \ref{sec:rnd} explores the nature of the DFT ground state as a function of U and lattice constant.    The appendix details how we calculated the magnetic quadrupole moments.

\section{Method}
\label{sec:method}

For our calculations we used the full potential, LAPW method within the local spin density approximation as implemented in Wien2k.\cite{wien2k}  Each of our calculations were performed on a $12 \times 12 \times 12 $ grid with 432 k-points in the irreducible Brillouin zone, a Fourier expansion of the potential up to a maximum reciprocal lattice magnitude of $G_{\mathrm{max}}=14$, and muffin tin radii of $R_{\mathrm{Mn}} = 2.197\ \mathrm{a.u.}$ and $R_{\mathrm{Si}} = 1.907\ \mathrm{a.u.}$.  The results did not vary with the size of the muffin tins up to a maximum of $R_{\mathrm{Mn}} = 2.3\ \mathrm{a.u.}$ and $R_{\mathrm{Si}} = 2.0\ \mathrm{a.u.}$.  We set the core/valence separation energy to include the Si 3s and 3p and Mn 3p, 3d, and 4s electrons in valence.  Also, we added local p and d orbitals to the Mn to eliminate ghost bands and improve convergence.  The convergence criteria that we used was that the integrated absolute difference between the electron densities on subsequent iterations was less than $0.001 \units{e}$.  Additionally, to ensure that our calculations were well converged with respect to energy, we varied the number of plane waves from $269$ to $1240$ by varying the \RmtKmax value between 5 and 9.  To this, we added a Hubbard-$U$ correction using the \AMF method to the 3d electrons on the manganese.\cite{lda+U.AMF.94}  

In the LDA+U \AMF method, the following is added to the LDA energy functional for each atom and electron shell for which additional correlations are desired,
\begin{equation}
	\label{eqn:AMF}
	\begin{split}
		E&^\text{AMF} = \frac{1}{2} \sum_{m, m\prime, \sigma} \Umat (n_{m \sigma} - \bar{n}_\sigma) 
							 (n_{m\prime,-\sigma} - \bar{n}_{-\sigma}) \\
						& + \frac{1}{2} \sum_{\substack{
								       				m,m\prime,\sigma \\
												m\ne m\prime}} (\Umat -\Jmat) (n_{m \sigma} - \bar{n}_\sigma) 
							(n_{m\prime \sigma} - \bar{n}_{\sigma}) 
	\end{split}
\end{equation}
where $n_{m \sigma}$ is the number of spin-$\sigma$ in the $m$ subband and $\bar{n}_\sigma$ is the average number of spin-$\sigma$ present in the entire shell.  The Coulomb, \Umat, and exchange, \Jmat, matrices are defined in terms of Slater integrals, $F^k$, and Clebsch-Gordan coefficients.  These are connected to the Coulomb repulsion, $U$, and exchange, $J$, energies by setting $U = F^0$, $J = (F^2 + F^4)/14$, and $F^4 = 0.625 F^2$.\cite{PhysRevB.48.16929}  We varied $U$ over the range $0.0\ \mathrm{Ry} \leq U \leq 0.7\ \mathrm{Ry}$, iterating to convergence at each value.  Like $U$, The exchange energy, $J$, is a fixed parameter of the system, and from optical spectra of Mn embedded in a metallic matrix, it is estimated to be $\sim 0.06 \units{Ry}$.\cite{harrison}  Constrained LDA calculations for MnO give a similar result.\cite{mno:08}  Our calculations show a small variation in the location of the transition region and the moment changes from $\mu \simeq 0.245$ to \mom{0.173}{Mn} as $J$ is varied from $0.04$ to $0.08 \units{Ry}$ at $U = 0.60 \units{Ry}$, and the results discussed in section \ref{sec:rnd} use the atomic value of $J=0.06 \units{Ry}$.

Additionally, we added a spin-orbit term to give the system a preferred spin direction, stabilizing the Hubbard-U calculation.  Initially, we chose [11\=2] and [10\=1] as our magnetization directions.  However, they gave similar results to the magnetization along [100] direction, which leaves the manganese ions equivalent, and is more computationally efficient.  The results discussed within are for the magnetization along the [100] direction.  Finally, we simulated hydrostatic pressure by varying the lattice constant between the experimental value of $4.558 \units{\AA}$ and $4.338 \units{\AA}$, equivalent to a maximum pressure of nearly $20 \units{GPa}$.\cite{MnSiPressure.exp.00}

\section{Results}
\label{sec:rnd}

\subsection{The Moment as a Function of $U$ and Pressure}
\label{sec:momwu}

\begin{figure}[htbp] 
  \centering
   \includegraphics[width=0.49 \textwidth]{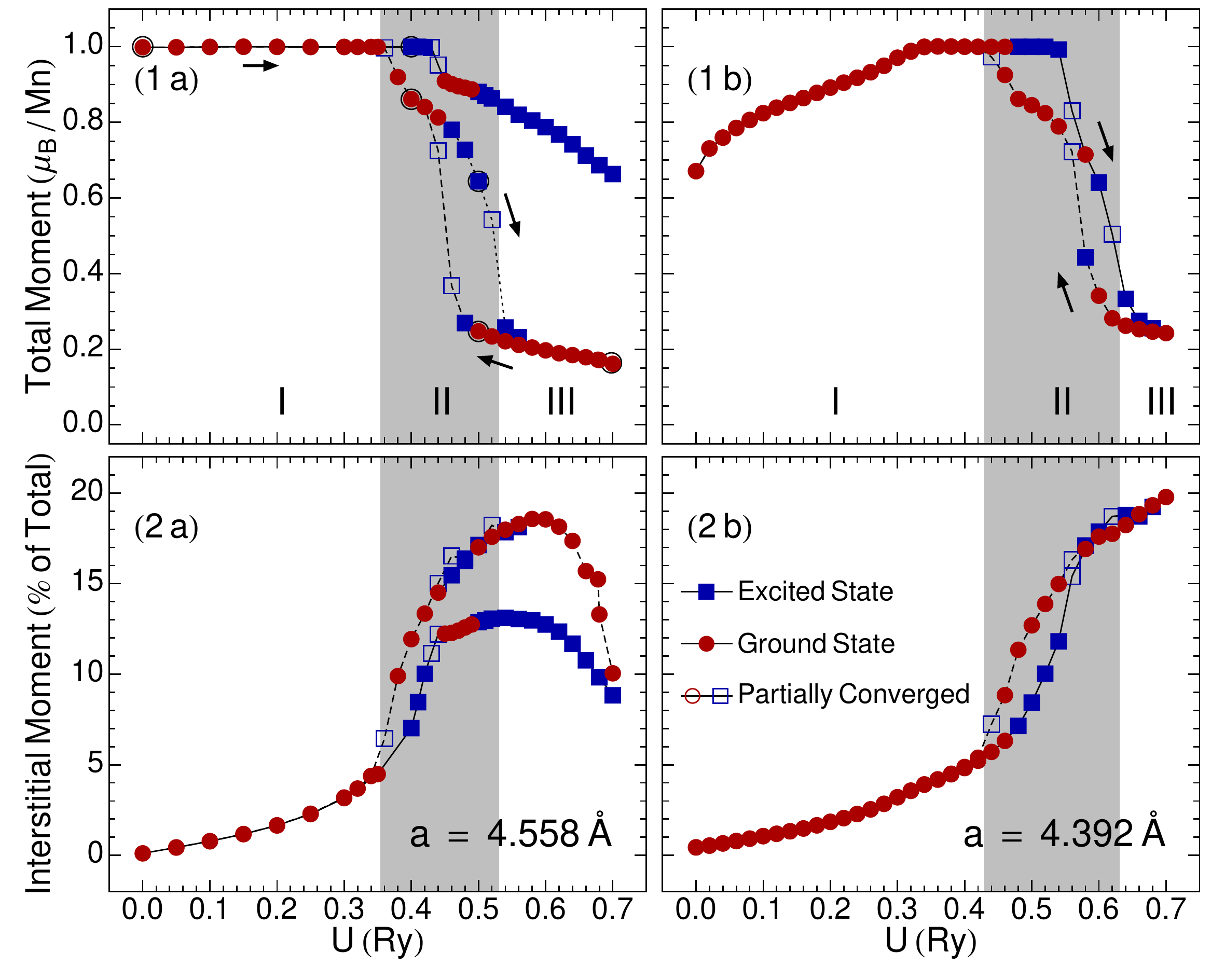}
    \caption{\small (color online) The total moment per manganese (row 1) and the percentage of the total moment in the interstitial regions (row 2) as a function of U at the both the experimental lattice constant ($4.558 \units{\AA}$, column a) and at high pressure ($4.398 \units{\AA}$, column b).  The ground (excited) state solutions are represented by a filled red circle (blue square), and partially converged solutions are only the outlines.  The three solution regions are labeled, and region II is highlighted.  The arrows are guides to the eye indicating how $U$ was changed along a particular branch.  There are several solutions circled in (1a) at the $U$ values of $0.0$, $0.4$, $0.5$, and $0.7 \units{Ry}$, and they are the ones used below in the subsequent table and figures. }
   \label{fig:data}
\end{figure}

In \figref{fig:data}, we show the total magnetic moment and the moment found in the interstitial regions of MnSi as a function of $U$ for two different lattice constants:  the experimental lattice constant at atmospheric pressure ($4.558 \units{\AA}$) and at high pressure ($4.398 \units{\AA}$) corresponding to about $14.8 \units{GPa}$.  At a number of values of U, we found multiple solutions.  We determined the ground state solutions by extrapolating from the total energy's variation with \RmtKmax to determine a best estimate total energy for each solution.  In a couple of cases, the ground state could not be determined within the error of the fit, so both solutions were listed as the ground state.  Additionally, the partially converged solutions are often in regions of function space where large rearrangements of charge and spin do not change the energy much, causing these solutions to require a large number of iterations to converge.

\begin{figure}[htbp] 
  \centering
   \includegraphics[width=0.49 \textwidth]{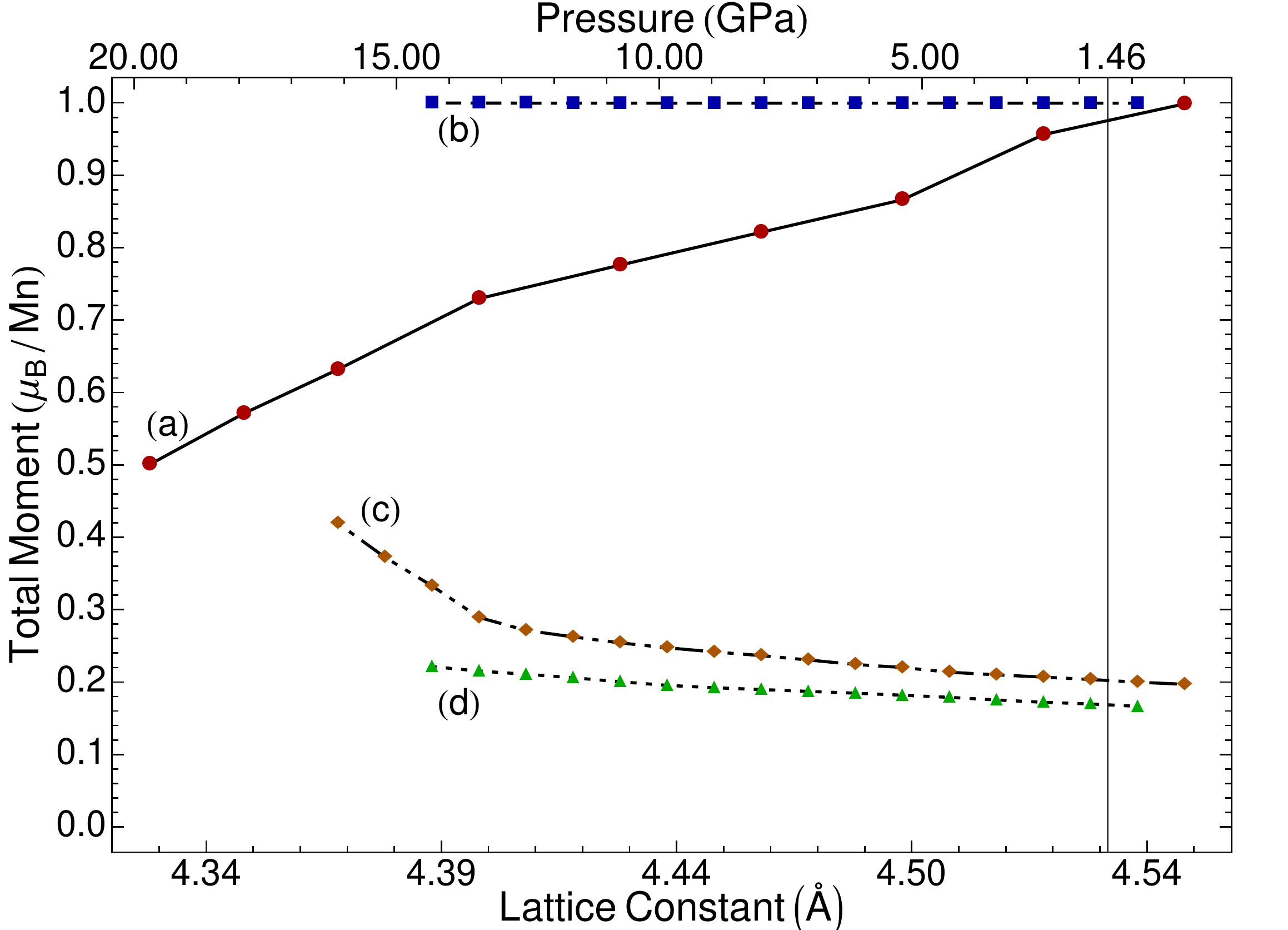}
    \caption{\small (color online) The total moment as a function of lattice constant for $U$ equal to  (a) $0.0 \units{Ry}$, (b) $0.4 \units{Ry}$ along the upper branch, and both (c) $0.6 \units{Ry}$ and (d) $0.7 \units{Ry}$ along the lower branch.  The upper axis is the pressure as determined by using the experimental bulk modulus.\cite{MnSiPressure.exp.00}  The gray vertical line is at the experimental metamagnetic transition point of $1.46 \units{GPa}$.}
   \label{fig:pressure}
\end{figure}

We have divided our solutions into three regions:  a high-moment, low-$U$ region (region I); a low-moment, high-$U$ region (III); and a complex transition between the two (II).  At the experimental lattice constant, we found a single ground state which maintains a moment of \mom{1}{Mn} in region I, and this moment is maintained into the transition region.  In general, the moments of the solutions in region I decrease with pressure, as shown in \figref{fig:pressure} (a) and (b).  But, this variation shrinks as $U$ is increased, until $U \approx 0.34 \units{Ry}$ where the moment does not change at the pressures we have looked at.  Additionally, we have been able to reproduce the work of \citeauthor{MnSiPressure.dft.99}, as  below a lattice constant of $\sim 4.336 \units{\AA}$ a non-magnetic solution becomes the ground state.\cite{MnSiPressure.dft.99}

Region I is also characterized by a highly localized spin density around the manganese atoms.  Most of the excess spin density within the lattice is found within a distance of less than half of the muffin tin radius from the Mn.  Additionally, as $U$ is increased the percentage of the total moment contributed by the interstitial regions grows from $\sim 0.1 \units{\%}$ to $\sim 5 \units{\%}$ at the region boundary.

Following the \mom{1}{Mn} moment solutions, we note that they quickly become metastable once region II has been entered, and the system switches to another solution with a lower moment as $U$ is increased.  At the experimental lattice constant, the system undergoes a transition to a state with a slightly reduced moment.  This state does not seem to be accessible at any other lattice constant, and we consider it spurious.  At higher pressure, the system undergoes a transition to a middle solution exhibiting hysteresis in $U$, and the moments of this branch are highly $U$ dependent.  With further increase in $U$, the system changes again to a lower moment solution that becomes the only accessible state in the third region.   As pressure is increased, the width of the region increases and shifts to the right.  Also in this region, the contribution from the interstitial moment increases at a faster rate, approaching $16 \units{\%}$ as region III is entered.

The third region is characterized by a single low moment state, with the exception of the spurious solution discussed above.  The states with slightly higher moments have converged within our tolerance level, and should be considered the same solution as they would have the same moment if a more stringent criteria were imposed.  The moments of the solutions in this region increase with pressure, as shown in \figref{fig:pressure} (c) and (d).  This is expected as the localizing effects of the Hubbard interaction are inversely proportional to the bandwidth, which is related to the interatomic spacing.\cite{Madelung}  So, as the pressure is increased, the system is pushed back towards the low-$U$ ground states.  Additionally, the interstitial contribution to the total moment remains high.  But, at the experimental lattice constant it peaks at $\sim 18 \units{\%}$ at $U \approx 0.6 \units{Ry}$, and decreases to $\sim 10 \units{\%}$ at $U = 0.7 \units{Ry}$. 

Additionally, we examined the behavior of the LDA minimum lattice constant with $U$, as outlined in \tblref{tbl:ldamin}.  Curiously, it behaves differently for the low-$U$ states and the high-$U$ states.  For the low-$U$ states, the LDA minimum is smaller than the experimental lattice constant and increase with $U$.  This is expected behavior as LDA alone underestimates bond lengths and the additional repulsive interaction would force the atoms farther apart.  However, the high-$U$ LDA minimum lattice constants are larger than the experimental value and they decrease with increasing $U$.  We believe this is caused by the decreasing spatial overlap between the majority and minority spin densities as $U$ is increased, which reduces the pressure caused by the on site interaction between the majority and minority spins.

\begin{table}
	\small
	\centering
	\begin{tabular}{r@{}l r@{}l r@{}l}
		\hline \hline
		$U$ & \units{(Ry)} & LDA M&in \units{(\AA)} & \% Diffe&rence  \\
		\hline
		0.&00 ({\small nonmagnetic}) & 4.&428 & -2.&85 \\
		0.&00  & 4.&438 & -2.&63 \\
		0.&40  & 4.&475 & -1.&82 \\
		0.&60  & 4.&675 & 2.&56   \\
		0.&70  & 4.&619 & 1.&34   \\
		\hline
	\end{tabular}
	\caption{\small The LDA minimum lattice constants were calculated and compared to the experimental lattice constant for the solutions shown in \figref{fig:pressure} and the nonmagnetic solution. }
	\label{tbl:ldamin}
\end{table}

\subsection{Density of States}

\begin{figure}[htbp] 
  \centering
   \includegraphics[width=0.485 \textwidth]{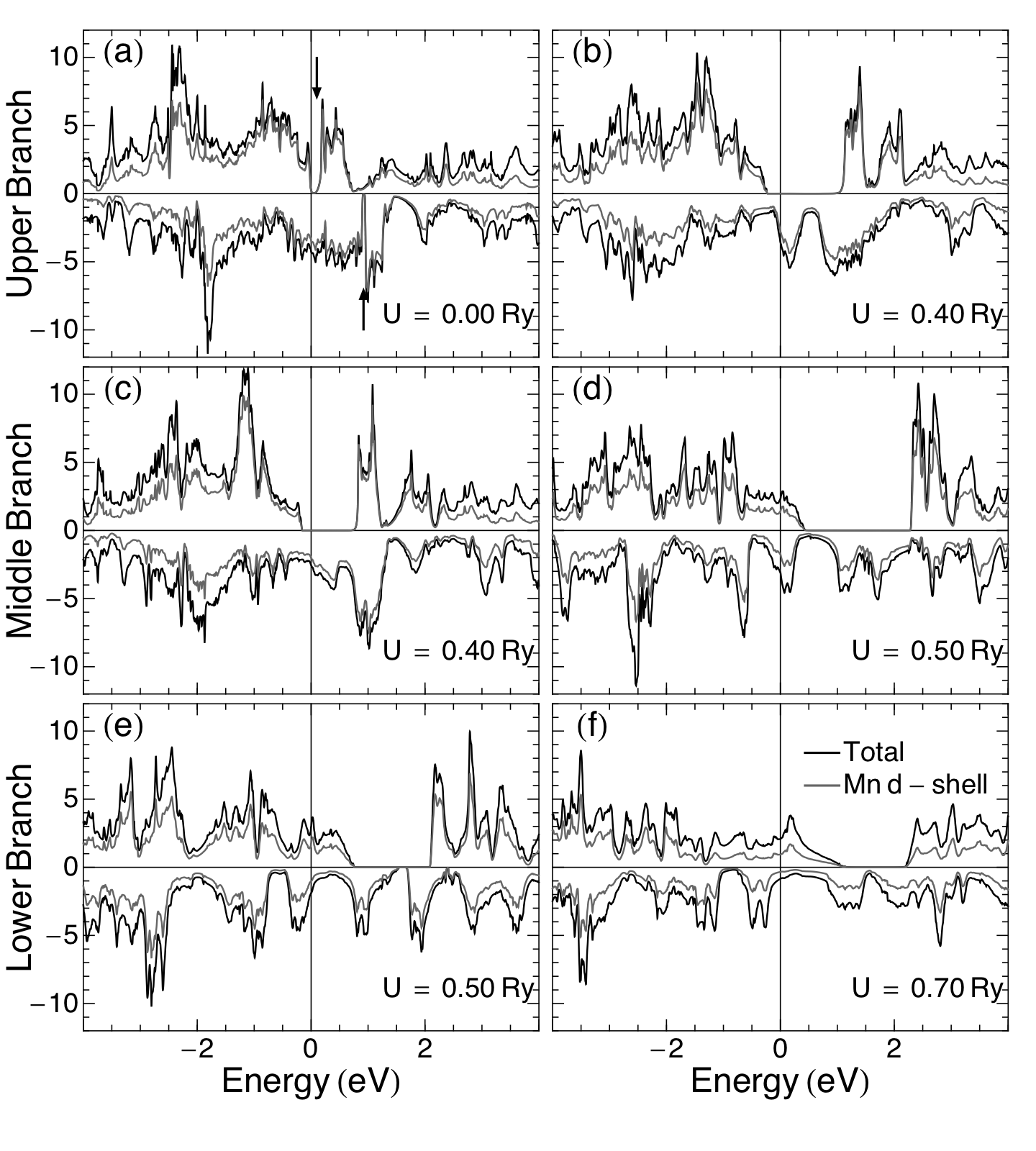}
    \caption{\small The total (black line) and manganese d-shell (grey line) spin resolved density of states for all six of the circled solutions in \figref{fig:data}(a) are shown.  The minority spin DOS is displayed along the negative vertical access.  In (a), there are arrows pointing out the majority and minority gaps.}
   \label{fig:dos}
\end{figure}

In order to understand the nature of the transition from the low-$U$ states to the high-$U$ states, we looked at the density of states (DOS), as shown in \figref{fig:dos}.  The most notable feature of our calculated DOS is the gap in the majority and minority spins.  At $U = 0 \units{Ry}$, there is a small gap in the minority spin, which disappears at the higher values of $U$.  However, the majority gap increases in size and shifts towards E$_F$ as $U$ increases.  The high-moment solutions become half-metallic at $U = 0.3 \units{Ry}$, and they remain so until the system transitions to another state.  For the low moment solutions, the gap remains large ($\gtrsim 1 \units{eV}$) and has jumped above the Fermi energy.  The second notable feature is the dramatic rearrangement the DOS undergoes as $U$ is varied.

For a single s-band, the addition of a Hubbard-$U$ splits the up and down states and forces them away from the Fermi energy.\cite{Mohn}  This essential feature is visible in the MnSi DOS as the d-shell states are forced away from each other as $U$ is changed from $0.0 \mbox{ to } 0.7 \units{Ry}$.  More precisely, at low values of $U$ the manganese d-shell states make up nearly all of states available near the Fermi energy.  But, as $U$ is increased, the d-shell electrons are forced away from E$_F$ flattening the peaks near E$_F$.  For $U = 0.5 \units{Ry}$ the Mn d states contribute about $70 \units{\%}$ of the total weight at E$_F$, and for $U = 0.7 \units{Ry}$ this is reduced to less than $50 \units{\%}$.   This contributes to the dramatic rearrangement of the states that occurs with the increase in $U$, and the DOS of the interstitial states grows to fill in the majority of the states available at E$_F$ for high values of $U$.  

\subsection{Angular Spin Distribution}

\begin{table}
	\begin{center}
	\subtable[]{
		\begin{tabular}{l r@{}l r@{}l  r@{}l r@{}l}
			\hline \hline
			State     \hfill \vline & 0.0 & \ (Ry) & 0.4 & \ (Ry) \hfill \vline  & 0.5 & \ (Ry)  & 0.7 & \ (Ry) \\
			\hline
			\dz       \hfill \vline & 0.&21 & 0.&22 \hfill \vline & 0.&18 & 0.&22 \\
			\dxyxy  \hfill \vline & 0.&40 & 0.&41 \hfill \vline & 0.&39 & 0.&41 \\
			\dxzyz  \hfill \vline & 0.&39 & 0.&37 \hfill \vline & 0.&43 & 0.&38 \\
			\hline
			Total     \hfill \vline & 3.&05 & 2.&99 \hfill \vline & 2.&62 & 2.&60   \\
			\hline
		\end{tabular}
	}
	\subtable[]{
		\begin{tabular}{l r@{}l r@{}l  r@{}l r@{}l}
			\hline \hline
			State     \hfill \vline & 0.0 & \ (Ry) & 0.4 & \ (Ry) \hfill \vline  & 0.5 & \ (Ry)  & 0.7 & \ (Ry) \\
			\hline
			\dz       \hfill \vline & 0.&18 & 0.&11 \hfill \vline & 0.&20 & 0.&13 \\
			\dxyxy  \hfill \vline & 0.&39 & 0.&37 \hfill \vline & 0.&40 & 0.&38 \\
			\dxzyz  \hfill \vline & 0.&43 & 0.&52 \hfill \vline & 0.&40 & 0.&49 \\
			\hline
			Total     \hfill \vline & 2.&03 & 2.&08 \hfill \vline & 2.&44 & 2.&46   \\
			\hline
		\end{tabular}
	}
	\end{center}
	\caption{\small The relative occupancies of the Mn d-subshells and the total occupancy of the Mn d-shells are shown for the majority (a) and minority (b) spins for several values of $U$.  The tables are divided into the low-$U$ and high-$U$ branches. \label{tbl:spin}}
\end{table}

\begin{figure}[htbp] 
  \centering
  \includegraphics[width=0.485 \textwidth]{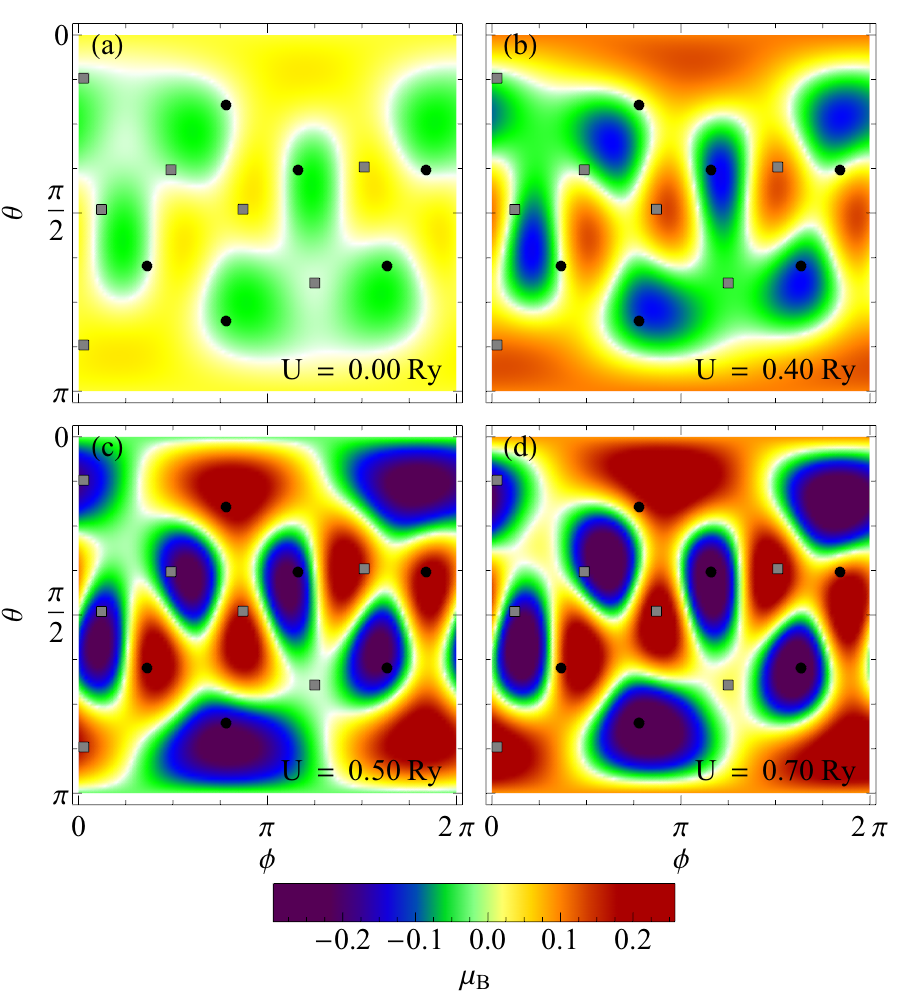}
  \caption{\small These are plots of the spin densities around the Mn atom radially integrated out to the muffin tin boundary for the circled solutions along the upper and lower branches in \figref{fig:data}(a).  The azimuthal angle, $\theta$, is plotted on the vertical axis and the polar angle, $\phi$, is plotted on the horizontal axis. The directions to the nearest neighbor Mn (Si) atoms are plotted as black circles (gray squares).}
   \label{fig:spin}
\end{figure}

\begin{figure}[htbp]
	\centering
	\subfigure[][$\, U=0.00\units{Ry}$]{\includegraphics[width=0.24 \textwidth]{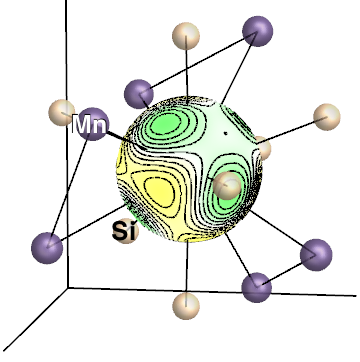}\label{fig:spin3d-u00}}%
	\subfigure[][$\, U=0.70\units{Ry}$]{\includegraphics[width=0.24 \textwidth]{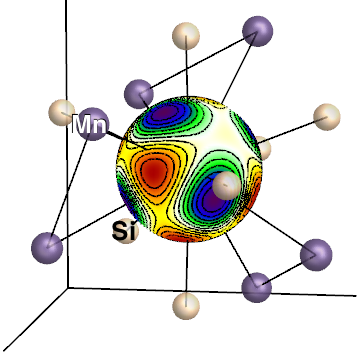}\label{fig:spin3d-u07}}	
	\caption{\small This is the same data as \figref{fig:spin} (a) and (d) plotted on the surface of a sphere the size of the Mn muffin-tin.  Included as a visual reference are the nearest neighbor atoms with lighter (darker) shading for the Si (Mn) atoms.  The bonding Si atom is labeled.}
	\label{fig:spin3d}
\end{figure}

By integrating the DOS up to E$_\text{F}$, we see that the decrease in the moment is caused by a small transfer of electrons from the majority to minority spin d-bands, as shown in \tblref{tbl:spin}.  This also reveals that the transfered electrons are not evenly distributed among the minority spin bands.  While the rearrangement is not large, it does cause a visible shift in the alignment of the spin density around the Mn atoms, as shown in Figs. \ref{fig:spin} and \ref{fig:spin3d}.

The most notable feature of the spin densities is the minority spin trilobal structures that are centered around the [111] direction.  Within in any branch of the solutions, the trilobal structure is distorted slightly as $U$ is varied.  However, the degree of spin polarization increases with increasing $U$ in the high moment branch (Figs. \ref{fig:spin}(a), (b) and \ref{fig:spin3d}).  In the case of the low-moment solutions, some regions have saturated and the boundary regions between the spin polarizations decrease in size.  Additionally, the center of the trilobal structures changes spin polarization as $U$ is increased in the low-moment branch.  Between the two solution branches, the trilobal structure is rotated about the [111] direction which tends to align the lobes with six of the thirteen nearest neighbors.  

This arrangement of spins is not unique to the having the magnetization aligned along the [100] direction.  With the magnetization aligned along the [10$\bar{1}$] direction at $U = 0.5 \units{Ry}$, the arrangement of the spin density is nearly the same as that shown in \figref{fig:spin}(c).  Since, the structure does not seem dependent upon the magnetization direction, nor do we find that it is affected by a $10 \units{T}$ field, the anisotropy may be detectible as a small magnetic quadrupole moment around the Mn atoms.

\subsection{Magnetic Quadrupole Moments}

\begin{figure}[htbp] 
  \centering
   \includegraphics[width=0.485 \textwidth]{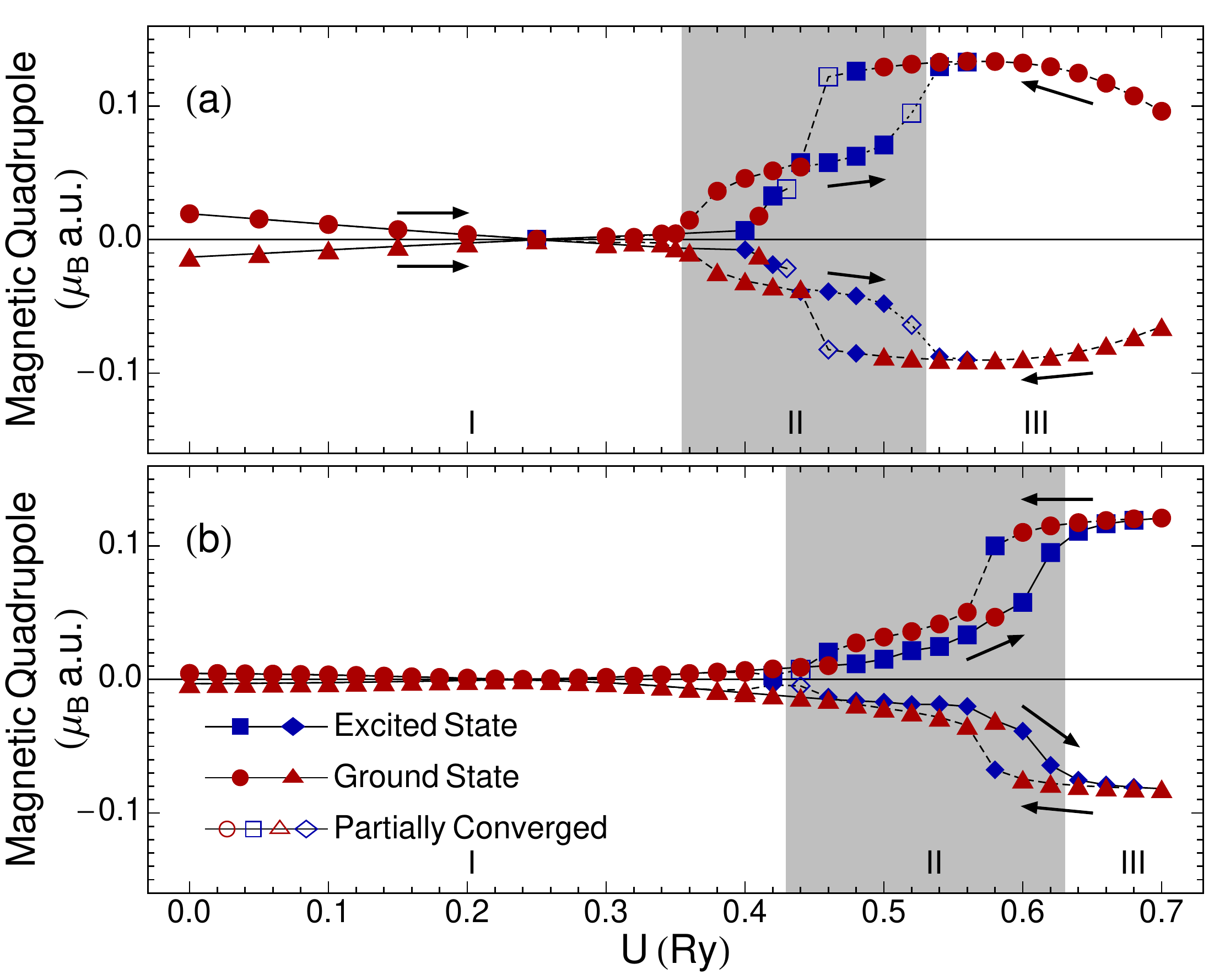}
    \caption{\small The two independent magnetic quadrupole moments as a function of $U$ for the lattice constants (a) $4.558 \units{\AA}$ and (b) $4.392 \units{\AA}$ are shown. The larger (smaller) moment is labeled by circles (triangles) and squares (diamonds) and lies above (below) the x-axis.  As in \figref{fig:data}, the three solution regimes are highlighted and labeled and the arrows serve the same purpose as before.  The quadrupole moments of the spurious solution are not shown.  }
   \label{fig:quad}
\end{figure}

From \linecite{magMultiPoles}, the magnetic quadrupole due to a non-uniform spin density can be calculated from the traceless tensor operator
\begin{equation}
	\mathcal{M}_{i j} = r_i s_j + r_j s_i - 2 \delta_{i j} r_k s_k
	\label{eqn:quadop}
\end{equation}
where $r_i$ and $s_i$ are the $i$th components of the position and spin operators, respectively.  In the appendix, we show this to be equivalent to the classical magnetic quadrupole.\cite{Jackson}  \figref{fig:quad} shows the two independent principal components of the magnetic quadrupole moments within the Mn muffin-tins for lattice constants $4.392 \mbox{ and } 4.558 \units{\AA}$ as a function of $U$.  

For the low-$U$ ground states, the magnetic quadrupole moments are negligible, and are further suppressed by the increase in pressure.  In the transition region, the quadrupole moments increase quickly, and show hysteresis with $U$ that was seen in the dipole moments.  The spurious solution at the experimental lattice constant, increases to the value obtained by the middle solution at $U=0.46 \units{Ry}$, but it does not increase beyond that point.  The middle solution quadrupole moments continue to increase as $U$ is increased.  The quadrupole moments are nearly their largest value of $0.15 \units{\mu_B a.u.}$ as the high-$U$ region is entered.  At the experimental lattice constant, the largest magnetic quadrupole is found at $\sim 0.58 \units{Ry}$, and it decreases with further increase in $U$.  Since an increase in pressure shifts the transition region to the right, the quadrupole moments continue to increase at least until $U=0.7 \units{Ry}$ for a lattice constant of $4.392 \units{\AA}$.  Additionally, the quadrupole moments around the silicon atoms does not exceed $0.03 \units{\mu_B a.u.}$, and the quadrupole moments in the nonmagnetic solution never exceeds a tenth of that value.

While these moments are small, the principal directions of the magnetic quadrupole are not aligned with the electric field gradient (EFG) at the Mn nucleus.  The angle between the eigenvectors for largest magnetic quadrupole moment and the principal direction of the EFG is about $27^\circ$ for $U$ between $0.5$ and $0.6 \units{Ry}$, and over that range, the orientation of the magnetic quadrupole stays nearly constant.  The principal direction of the EFG, also, remains roughly constant over that range, but the other directions are rotated $\sim 24^\circ$ about the principal direction.  It should be noted, that the principal directions of the magnetic quadrupole moment are mostly insensitive to pressure, and while the principal direction of the EFG is also insensitive to pressure, the other EFG directions rotate about $21^\circ$ from their original positions.  Since both M\"ossbauer spectroscopy\cite{PhysRevB.30.32} and $^{55}$Mn NMR\cite{JPSJ.44.833} are capable of detecting anisotropies in the hyperfine field, they should be able to detect the magnetic quadrupole moment.

\section{Discussion}
\label{sec:conc}

Empirically, $U$ was found to be $0.13 \units{Ry}$, assuming that $J = 0.06 \units{Ry}$ and neglecting the multipole contribution.\cite{carbone:085114}  At this value of $U$, the moment remains \mom{1}{Mn} and it is less sensitive to pressure, so this value of $U$ is clearly unsuitable.  In an LDA calculation for MnO, a value of $U = 0.51 \units{Ry}$ was found.\cite{mno:08}  The ground state at the computed value lies on the lower branch within the transition region with a moment of \nmom{0.24}{Mn}.  While this solution has a moment that is comparable to experiment, the hyperfine field around the Mn nucleus is $H_\text{HF} = 13 \units{kG}$ when the orbital and dipolar contributions are included.  However, increasing $U$ to $0.62 \units{Ry}$ gives $H_\text{HF} = -2.1 \units{kG}$, which is in good agreement with experiment.\cite{PhysRevB.42.6515}  The hyperfine field at the Si nucleus, however, is about half the measured $-91.3 \units{kOe}$ and aligned parallel with the magnetic moment,\cite{corti:115111} but the moment is only \mom{0.19}{Mn}.  As for the pressure dependence of these low moment states, we know that the experimental high pressure state exhibits a complex magnetic order.\cite{non-fermi:01}  Since the magnetic moment is well reproduced with LDA+U, this suggests that dynamic correlations are responsible for the observed pressure dependence.

In summary, we have found a set of solutions with a moment comparable to experiment for values of $U$ that are in line with prior work on other Mn compounds.  These solutions exhibit a distinct magnetic anisotropy that manifests as a small magnetic quadrupole moment around the Mn, which should be experimentally detectable.  Additionally, they correctly determine the hyperfine field at the Mn atom.  However, these solutions behave incorrectly under pressure and predict the wrong hyperfine field at the Si atom.

\begin{acknowledgments}
We thank John DiTusa and Richard Kurtz for useful conversations. This work is supported in part by NSF-OCI Grant No. 043812 and a Faculty Research Grant from the LSU Office of Sponsored Research. 
\end{acknowledgments}

\appendix*

\section{Calculating Magnetic Quadrupole Moments}
\label{sec:spleen}
According to \linecite{Jackson}, the multipole moments of the magnetization density, $\mathbf{M}(\mathbf{x})$, can be calculated by the formulations for the moments of the charge density by substituting $\rho_\text{M} = -\nabla \cdot \mathbf{M}$ for the charge density.  The cartesian form of the magnetic quadrupole is then
\begin{equation}
	\label{eqn:xfquad}
	\mathcal{Q}^{(\text{M})}_{i j} = -\frac{1}{6}\int(3 x_i x_j - \delta_{i j} x_l  x_l) \partial_k M_k\, \ud^{3}x,
\end{equation}
where the summation convention has been employed and $\partial_i = \partial/\partial x_i$.  To connect this formula to the operator in \eqnref{eqn:quadop}, we must integrate by parts.   By first noting that $\partial_i x_j = \delta_{i j}$,  the first term in the integrand becomes
\begin{equation}
	\label{eqn:fst}
	x_i x_j \partial_k M_k = \partial_k (x_i x_j M_k) - (x_i M_j + x_j M_i)
\end{equation}
and the second term is
\begin{equation}
	\label{eqn:snd}
	\delta_{i j} x_l x_l \partial_k M_k = \delta_{i j} [ \partial_k (x_l x_l M_k) - 2 x_l M_l ].
\end{equation}
Substituting these into \eqnref{eqn:xfquad}, we get
\begin{align}
	\label{eqn:fullquad}
	\mathcal{Q}^{(\text{M})}_{i j} = &- \frac{1}{6} \int \partial_k (x_i x_j M_k - \delta_{i j} x_l x_l M_k) \, \ud^{3}x
	\nonumber\\ &+\frac{1}{6} \int [ 3 (x_i M_j + x_j M_i) - 2 \delta_{i j} x_k M_k] \, \ud^{3}x
\end{align}
The first term is the surface term, and in the case of a localized density it can be taken to be zero.\cite{Jackson}  Since we are integrating over the Mn muffin tins, the density is not localized and there are spurious monopole terms if the surface terms are ignored.  The second term is the expectation value of the magnetic quadrupole operator in \eqnref{eqn:quadop}.  This can be seen by noting that the spin density, $M_z = \rho_\uparrow - \rho_\downarrow$, is related to the ground state spinor, $\mathbf{\Psi}$, by $\mathbf{\Psi}^\dagger \cdot \sigma_z \cdot \mathbf{\Psi}$ where $\sigma_z$ is the Pauli matrix.  In our calculations, we use expectation value of $\mathcal{M}$ directly, as it contains the surface term.

\bibliography{../MnSi.bib,../generalMaterial.bib,../electronicStructure.bib,../transport.bib}

\end{document}